\documentclass[%
 reprint,
 amsmath,amssymb,
 aps,
 prb,
 floatfix,
]{revtex4-1}

\usepackage{graphicx}
\usepackage{dcolumn}
\usepackage{bm}
\usepackage[colorlinks=true, linkcolor=blue, citecolor=blue, urlcolor=blue]{hyperref}
\usepackage{stmaryrd}
\usepackage{txfonts}
\usepackage{mathrsfs}
\usepackage{dcolumn}
\usepackage{braket}
\usepackage{color}
\usepackage[normalem]{ulem}
\usepackage{bibunits}
\usepackage{xspace}
\usepackage{accents}
\newcommand{\Tsf}{T_\mathrm{SF}} 
\newcommand{\Tsdw}{T_\mathrm{SDW}}
\newcommand{\Tli}{T_\mathrm{IC-C}}
\newcommand{\Qcom}{\mathbf{q}_\text{com}}
\newcommand{\Qmin}{\mathbf{q}_\text{min}}
\newcommand{\field}{\psi}
\newcommand{\SDW}{\phi_c}
\newcommand{\mzero}{m}
\newcommand{\GLH}{\mathscr{H}}

\begin{document}

\title{
    Magnetic field-induced deformation of the spin-density wave microphases in Ca$_3$Co$_2$O$_6$
}

\author{
    Y. Kamiya
}
\affiliation{
    School of Physics and Astronomy, Shanghai Jiao Tong University, Shanghai 200240, China
}
\date{\today}
             
\begin{abstract}
The frustrated triangular Ising magnet Ca$_3$Co$_2$O$_6$ has long been known for an intriguing combination of extremely slow spin dynamics and peculiar magnetic orders, such as the evenly-spaced non-equilibrium metamagnetic magnetization steps and the long-wavelength spin density wave (SDW) order, the latter of which is essentially an emergent crystal of solitons. Recently, an elaborate field-cooling protocol to bypass the low-field SDW phase was proposed to overcome the extraordinarily long timescale of spin relaxation that impeded previous experimental studies in equilibrium, which may point to a deep connection between the low-temperature slow relaxation and the cooling process passing through the low-field SDW phase. As the first step to elucidate the conjectured connection, we investigate the magnetic field-induced deformation of the SDW state and incommensurate-commensurate transitions, thereby mapping out the equilibrium in-field phase diagram for a realistic three-dimensional lattice spin model by using Monte Carlo simulations. We also discuss Ginzburg-Landau theory that includes several Umklapp terms as well as an effective sine-Gordon model, which can qualitatively explain the observed magnetic field-induced deformation of the SDW microphases.
\end{abstract}

                              
\maketitle

\section{
Introduction
\label{sec:introduction}
}

Frustrated magnets can have a manifold of nearly degenerate low-energy states from which interesting phenomena may emerge, such as exotic magnetic and nonmagnetic orders, topological order, liquid-like or even glassy behavior, and so on, varying from one material to another.\cite{Lacroix2011} Even a classical system can host unconventional quasiparticles, such as skyrmions,\cite{Bogdanov2020} solitons, and monopoles in spin ice,\cite{Bramwell01,Paulsen2014,Bramwell20} and they may crystallize into novel spin textures like skyrmion crystals~\cite{Okubo2012,Leonov2015} or soliton crystals.\cite{Bak82,Selke88} Such emergent crystalline states can often be sensitive to external perturbations, which makes them attractive as potential devices in some cases.\cite{Fert2017} They can also provide a platform to study far-from-equilibrium dynamics due to metastable states.\cite{Kudasov06,Paulsen2014}

\begin{figure}[b]
\begin{center}
    \includegraphics[width=\columnwidth]{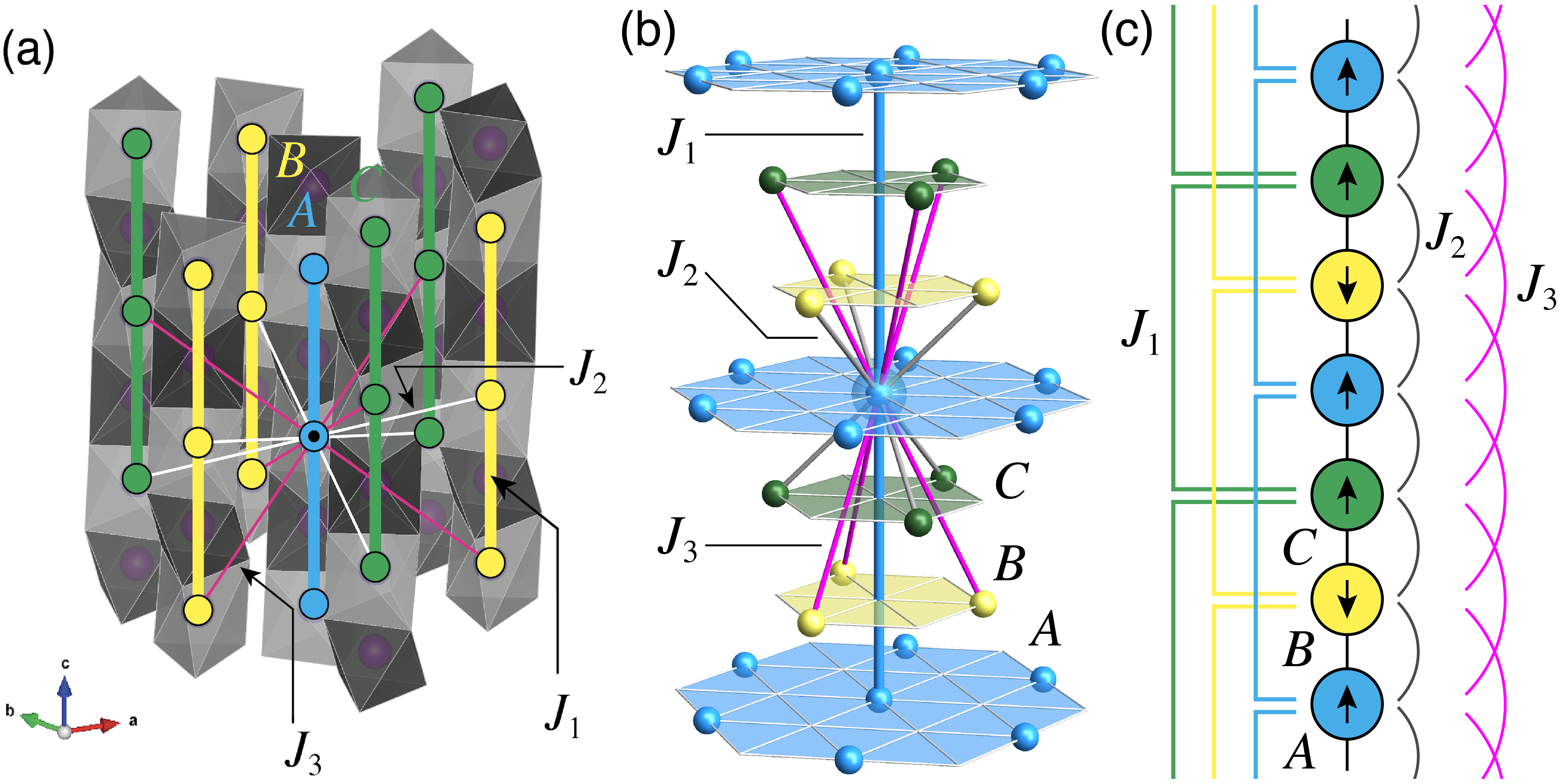} 
\end{center}
\caption{
    \label{fig:lattice}
    (a) Magnetic lattice of CCO with the intrachain coupling $J_1$ and the interchain couplings $J_2$ and $J_3$, which is superimposed on the schematic crystal structure with sublattices $A$, $B$, and $C$. 
    (b) Another schematic picture for the magnetic lattice, where only those exchange interactions that connect to the central site are shown to illustrate them in a concise manner.
    (c) Effective one-dimensional lattice used in the mean field theory (see the text), where an arrow represents a magnetic moment of a given $ab$ plane.
}
\end{figure}

Since the late 90's,\cite{Fjellvag96,Aasland97,Kageyama97,Kageyama97b} the frustrated triangular Ising magnet Ca$_3$Co$_2$O$_6$ (CCO) has long been known for an intriguing combination of extremely slow spin dynamics and peculiar magnetic phases, such as metamagnetic magnetization steps and an incommensurate spin-density wave (SDW) state,\cite{Agrestini08a,Mazzoli09,Agrestini08b,Moyoshi11,Fleck10,Paddison14,Motoya18} the latter of which can be seen as a soliton crystal.\cite{Kamiya12} In CCO, trigonal prismatic Co$^{3+}$ $S = 2$ sites form ferromagnetic Ising chains running along the $c$ axis, while arranged in a triangular lattice in the $ab$ plane coupled by weak antiferromagnetic interactions (Fig.~\ref{fig:lattice}).\cite{Leedahl19} Below spin-freezing temperature $\Tsf \simeq 5\,\mathrm{K}$, CCO exhibits striking evenly-spaced metamagnetic magnetization steps,\cite{Kageyama97} whose origin has been a subject of long-time debates.\cite{Kageyama97,Kageyama97b,Maignan00,Hardy04b,Maignan04,Moyoshi11,Kim18} Interestingly, while the step heights are sensitive to protocol details such as sweep rate of the external magnetic field, the transition magnetic fields ($\simeq$1.2\,T, 2.4\,T, and 3.6\,T, with additional steps at higher fields) are rather robust.\cite{Hardy04b} Although some theory invoked an analogy with quantum tunneling in molecular magnets,\cite{Maignan04} an alternative scenario is that peculiar frustration in CCO causes a non-equilibrium phenomenon.\cite{Kudasov06} In the so-called ``rigid chain'' model,\cite{Yao06a,Yao06b,Kudasov08,Qin09,Soto09,Zukovic12} each ferromagnetic chain is replaced by an effective Ising spin on a two-dimensional antiferromagnetic triangular lattice. Based on this mapping, it was argued that the metamagnetic transition steps in CCO may arise from the same kind of degenerate manifold as in the two-dimensional triangular lattice Ising model.\cite{Kudasov06} 

However, the origin of the slow dynamics can be more intricate than suggested by the rigid chain picture. More recently, resonant x-ray~\cite{Agrestini08a,Mazzoli09} and neutron spectroscopies~\cite{Agrestini08b,Moyoshi11,Fleck10,Paddison14,Motoya18} revealed the SDW order below $\Tsdw \simeq$ 25~K, which has a three-sublattice structure and a very long modulation wavelength $\lambda_\mathrm{SDW}\!\simeq\! 10^3$\,$\AA$ ($\simeq\!\!10^2$ magnetic sites) along the $c$ axis. It was found that $\lambda_\mathrm{SDW}$ increases as temperature $T$ is lowered and the corresponding relaxation time grows substantially. Eventually, the system starts to deviate from equilibrium below $T \lesssim$ 13~K,\cite{Moyoshi11} which is much higher than $\Tsf$ for the appearance of the metamagnetic magnetization steps. Since the spin chains are not ferromagnetically ordered in the SDW state, the interpretation of the rigid chain picture is, if not questionable, more subtle than originally proposed.

Indeed, the SDW phase may hold the key to understanding the peculiar slow dynamics at low temperatures. It was recently demonstrated that the slow spin dynamics can be bypassed by an elaborate field-cooling protocol, where every in-field measurement is performed after a separate cooling in the target magnetic field.\cite{Nekrashevic21} Remarkably, the protocol allowed for reaching the 1/3 magnetization plateau down to $T = 2\,\mathrm{K} < \Tsf$ without being suffered from metastable states, which was in good agreement with MC simulations in equilibrium. Since the SDW order is believed to disappear and replaced by a ferrimagnetic state at high magnetic fields, the new experiment may suggest that the spin relaxation at low temperatures may be influenced by the extent to which the system has been through the low-field SDW phase during the cooling. In fact, it is known that the SDW order is accompanied by short-range order indicative of spin disordering,\cite{Agrestini11} which could be due to the combination of the $T$-dependent ordering wavevector and the slowness in the relaxation to follow the variation.~\cite{Moyoshi11}

As mentioned above, the observed SDW state is essentially a soliton crystal as in the axial next-nearest-neighbor Ising (ANNNI) model,\cite{Kamiya12} a prototypical model for spontaneous superstructures due to competition between nearest and next nearest neighbor Ising interactions in one direction of a square or cubic lattice.\cite{Bak82,Selke88} The $T$-dependent change of $\lambda_\mathrm{SDW}$~\cite{Moyoshi11} corresponds to different magnetic \emph{microphases},\cite{Kamiya12} similar to other self-organizing modulated phases in physical and chemical systems.\cite{Seul1995} Thus, CCO may provide a rare intersection where the ANNNI model phenomenology~\cite{Bak82,Selke88} meets out-of-equilibrium physics in a solid state system with only short-range interactions. To elucidate this conjecture in CCO and similar materials such as Ca$_3$Co$_{2-x}$Mn$_x$O$_6$,\cite{Zapf18,Kim18} Sr$_2$Ca$_2$CoMn$_2$O$_9$,\cite{Hardy18} and Ca$_3$CoRhO$_6$,\cite{Niitaka01} it is a matter of paramount importance to investigate the magnetic field-induced deformation of the SDW state and incommensurate-commensurate (IC-C) transitions under the condition much closer to equilibrium than ever reached before. The goal of this work to present an equilibrium in-field phase diagram for a realistic three-dimensional (3D) lattice spin model for CCO, thereby providing a theoretical guide for experiments. We address both model-specific and universal physics by combining mean-field theory (Sec.~\ref{Sec:MF}), MC simulations (Sec.~\ref{Sec:MC}), and Ginzburg-Landau (GL) theory (Sec.~\ref{Sec:GL}).

\section{
Model
\label{Sec:model}
}

In CCO, $S = 2$ spins have large easy-axis anisotropy,\cite{Kageyama97} which permits a description by an effective classical Ising model, 
\begin{align}
    \hat{H} = \sum_{\nu = 1,2,3}\sum_{\langle{ij}\rangle_{\nu}} J_\nu \sigma^z_i \sigma^z_j - h \sum_i \sigma^z_i,
    \label{eq:H}
\end{align}
where $\sigma^z_{i} = \pm 1$, $h = g \mu_B S H$ with $g$, $\mu_B$, and $H$ being the $g$-factor, the Bohr magneton, and a magnetic field, respectively, and $\langle{ij}\rangle_{\nu}$ denotes neighboring sites connected by $J_\nu$, $\nu \in \{1,2,3\}$. $J_1 < 0$ is the intrachain ferromagnetic interaction and $J_2$ ($J_3$) is the antiferromagnetic interchain interaction shifted by 1/3 (2/3) lattice parameters along the $c$ axis (Fig.~\ref{fig:lattice}). An \textit{ab initio} study suggested $\lvert{J_1}\rvert \gg J_2 \simeq J_3$~\cite{Fresard04} and $J_1 = -23.9(2)$\,K and $J_2 + J_3 = 2.3(2)$\,K was reported by an NMR experiment, which further suggested $J_2 = 1.1$\,K and $J_3 = 1.2$\,K to explain the SDW ordering wavevector.\cite{Allodi14} Below, for simplicity, we assume $J_2 /\lvert{J_1}\rvert = J_3 /\lvert{J_1}\rvert$ and denote the ratio by $\kappa$; in relation with CCO, $\kappa \simeq 0.048$.

\begin{figure}[t]
\begin{center}
    \includegraphics[width=0.9\columnwidth]{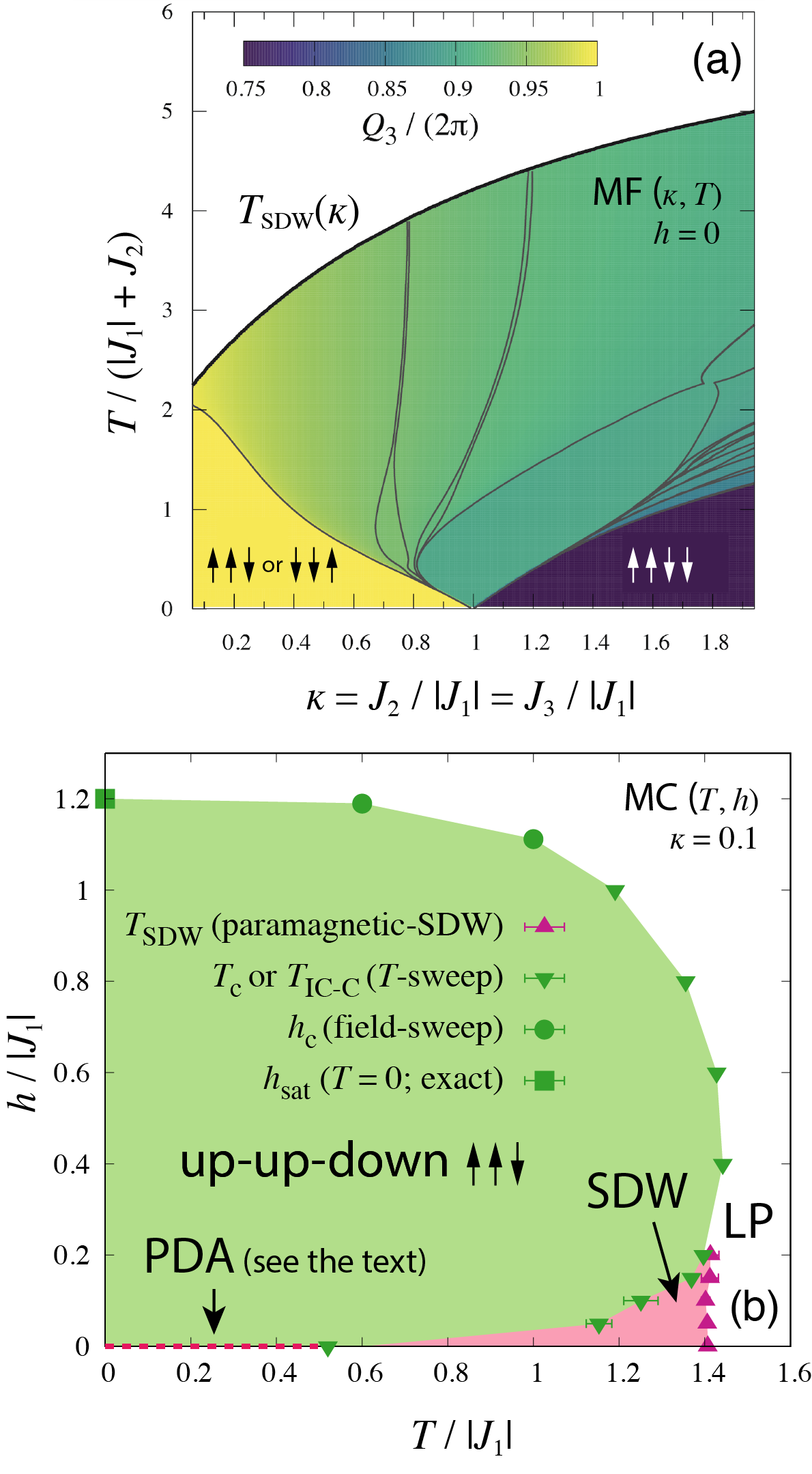}
\end{center}
\caption{
 \label{fig:phase_diagram}
    (a) Mean field $(\kappa, T)$ phase diagram  for $h=0$, where $\kappa \equiv J_2 / \lvert{J_1}\rvert = J_3 / \lvert{J_1}\rvert$. The mean field ground state is $\uparrow\uparrow\downarrow$ or $\downarrow\downarrow\uparrow$ ($\uparrow\uparrow\downarrow\downarrow$) along the $c$ axis for $\kappa < 1$ ($\kappa > 1$) in the effective one-dimensional description, the ordering wavevector of which is $Q_3 / (2\pi) = 1$ ($3/4$), respectively. 
    (b) ($T$, $h$) phase diagram obtained by MC simulations for $\kappa = 0.1$ based on the data for $T \gtrsim 0.45 \lvert{J_1}\rvert$. The $T$-sweep data for $T_c$ or $\Tli$ is determined by analyzing the Binder parameter as a function of $T$, whereas the field-sweep data for $h_c$ is determined from a peak in $\chi = dM/dh$ as a function of $h$.
}
\end{figure}

\begin{figure*}
\begin{center}
    \includegraphics[width=\hsize]{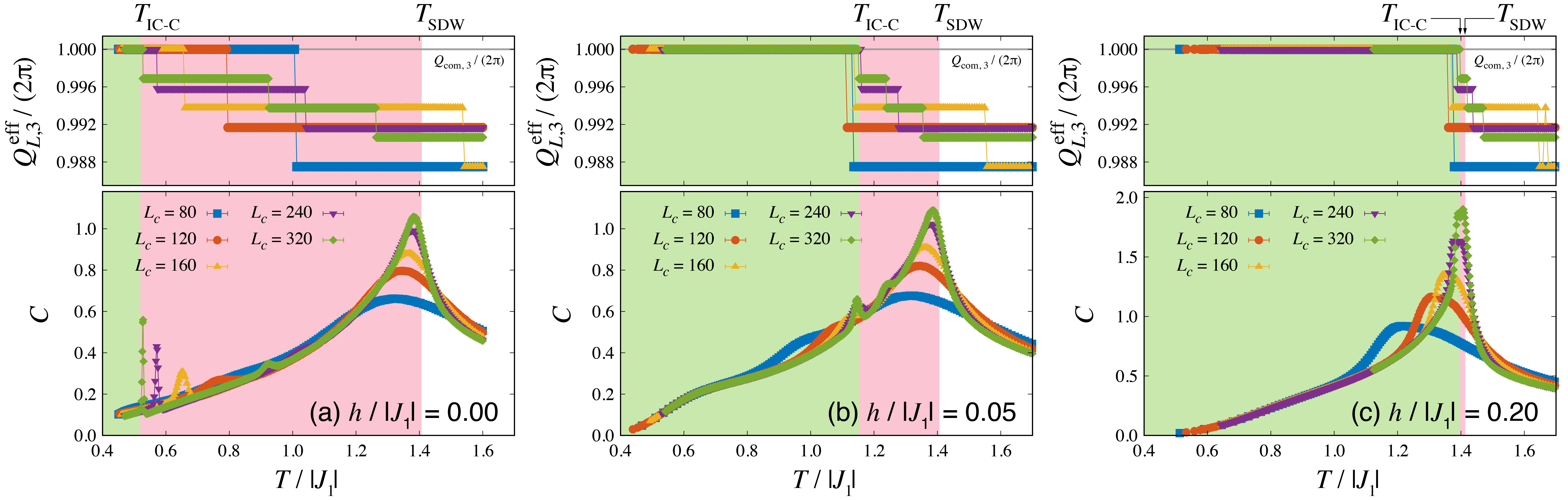}
\end{center}
\caption{
 \label{fig:C_and_Q}
 $T$-dependence of the third component of the effective ordering wavevector $Q^\mathrm{eff}_{L,3}$ (upper panels) and specific heat $C$ (lower panels) obtained by MC simulations for $\kappa = 0.1$ and (a) $h = 0$, (b) $h / \lvert{J_1}\rvert = 0.05$, and (c) $h / \lvert{J_1}\rvert = 0.2$. The additional subpeaks in $C$ are considered as spurious ones (see the text).
}
\end{figure*}

\section{
Mean field theory
\label{Sec:MF}
}
In CCO, $J_2$ and $J_3$ compete with $J_1$ after a few steps along a spiral path due to the vertical shifts of the interchain interactions.\cite{Chapon09} This spiral structure is the key to realize the same kind of geometrical frustration as in the ANNNI model~\cite{Bak82,Selke88} despite the apparent structural differences. We first briefly discuss a heuristic mean-field theory in zero field by assuming a ferromagnetic order in each $ab$ plane, which are separated by 1/3 lattice parameters from each other along the $c$ axis (Fig.~\ref{fig:lattice}).\cite{Kamiya12} The mean field equation for the magnetization $m_l$ of layer $l$ is
\begin{align}
    \left\langle{m_l}\right\rangle = \tanh \beta h_l,
\end{align}
where $h_l = -J_1(\left\langle{m_{l+3}}\right\rangle + \left\langle{m_{l-3}}\right\rangle) - 3J_2(\left\langle{m_{l+1}}\right\rangle + \left\langle{m_{l-1}}\right\rangle) - 3J_3(\left\langle{m_{l+2}}\right\rangle + \left\langle{m_{l-2}}\right\rangle)$. In this quasi-one-dimensional description, $J_1$, $J_2$, and $J_3$ serve as the effective third, first, and second neighbor interactions, respectively, realizing a very similar situation as in the prototypical ANNNI model [Fig.~\ref{fig:lattice}(c)]. The reason for assuming an in-plane ferromagnetic order, even though the interchain interactions $J_2$ and $J_3$ are much smaller than the intrachain interaction $J_1$, is that the energy scale associated with the competition between SDW states with different wavelengths along the $c$ axis can be even smaller, as will be discussed by using a sine-Gordon model. In Fig.~\ref{fig:phase_diagram}(a), we show the mean field $(\kappa,T)$ phase diagram, by extending the previous work.\cite{Kamiya12} Below the SDW transition temperature $\Tsdw(\kappa)$, the ordering wavevector $\mathbf{Q} = (0,0,Q_3)$ varies quasi-continuously. Eventually, there is a lock-in IC-C transition at $T = \Tli$, below which the magnetic unit cell of the mean field solution is $\uparrow\uparrow\downarrow$ or $\downarrow\downarrow\uparrow$ for $\kappa < 1$ and $\uparrow\uparrow\downarrow\downarrow$ for $\kappa > 1$ in the effective one-dimensional description in Fig.~\ref{fig:lattice}(c). We find that quasi-continuous changes of $Q_3(\kappa,T)$ dominate the overall phase diagram, corresponding to numerous  microphases of soliton lattice states, especially for relatively small $\kappa$.\cite{Kamiya12} Meanwhile, distinct discontinuous changes of $Q_3(\kappa,T)$ are also seen in the region with relatively large $\kappa$, where a few relatively extended commensurate states are found. However, the latter case has little significance in relation with CCO, where $\kappa$ has been suggested to be very small.

\section{
Monte Carlo simulation
\label{Sec:MC}
}
\subsection{
Set up
}

Next, to demonstrate the ANNNI-like physics in an unbiased way, we present the results of our MC simulations. We consider a lattice of size $L \times L \times L_c$ with periodic boundary conditions, where the total number of spins is $N_\text{spin} = 3L^2 L_c$. We combine single-spin updates, intra-chain cluster  updates~\cite{Kamiya12}, and replica exchanges~\cite{Hukushima96} included every 10 MC steps. Several hundreds of replicas are needed for largest lattices to maintain a reasonable exchange acceptance rate to guarantee efficient sampling at low temperatures (e.g., 400 replicas for $8\times 8\times 320$ for $h \lesssim 0.2\lvert{J_1}\rvert$). By fixing $\kappa = 0.1$ and $L_c / L = 40$ in most cases shown below, we performed simulations for $2\times 2 \times 80$--$8\times 8 \times 320$. Here, although $\kappa = 0.1$ is larger than $\kappa \simeq 0.048$ estimated for CCO, no qualitatively different physics for smaller $\kappa$ is suggested in our mean field phase diagram, as long as the SDW order is concerned [Fig.~\ref{fig:phase_diagram}(a)].

The aspect ratio $L_c / L = 40$ is chosen to simulate long-wavelength SDW states with as little finite-size tension as possible while not making the system excessively anisotropic to address thermodynamic behaviors in 3D. As discussed by using a GL theory, the ordering wavevector $\mathbf{Q}$ at $T = \Tsdw$ is expected to be the minima $\pm\Qmin$ of the Fourier transform $J(\mathbf{q})$ of the exchange interactions, which we find $\Qmin = (0,0,2\pi + \epsilon)$ with $\epsilon \approx -0.006 \times (2\pi)$ for $\kappa \simeq 0.048$ and $\epsilon \approx -0.013 \times (2\pi)$ for $\kappa = 0.1$. Because $\Qmin$ is very close to the three-sublattice commensurate wavevector $\Qcom = (0,0,2\pi)$, even a single periodicity of the spin modulation requires a large number of unit cells along the $c$ axis. (Here, $\mathbf{q} = 3\Qcom = (0,0,6\pi)$ is equivalent to $\mathbf{q} = 0$ in our notation, but $\mathbf{q} = \Qcom$ is not.) The minimum size thus required for $\kappa \simeq 0.048$ is $L_c^\text{min} = 2\pi / \lvert{q_\mathrm{min,3} - q_\mathrm{com,3}}\rvert \approx 160$, which can be a bit problematic. For $\kappa = 0.1$, we find $L_c^\text{min} \approx 80$, which is also quite anisotropic but within the acceptable range. The aspect ratio $L_c / L = 40$ is determined on this basis. 

\subsection{
Modified Binder parameter method
}

To determine the transition point of a second-order phase transition into a commensurate ordering, say with a wavevector $\mathbf{Q}$, a standard method is to analyze the order parameter $M_{\mathbf{Q}} = N_\text{spin}^{-1} \sum_{i} \sigma^z_i \exp(-i\mathbf{Q}^{}\cdot \mathbf{r}^{}_i)$ and the corresponding Binder parameter $U^{}_{\mathbf{Q}} = \langle{\lvert{M_{\mathbf{Q}}}\rvert^4}\rangle / \langle{\lvert{M_{\mathbf{Q}}}\rvert^2}\rangle^2$.\cite{LandauBinder} However, the possible $T$-dependent variation as well as incommensurability of the ordering wavevector poses a challenge in numerical studies of the ANNNI model and its variants,\cite{Gendiar2005,KaiZhang2010,KaiZhang2011} demanding a modified approach. In such a model, a finite-size system with periodic boundary conditions is expected to develop a spin correlation whose dominant wavevector is necessarily commensurate but near the true ordering wavevector $\mathbf{Q}(T)$ in thermodynamic limit. Such an effective ordering wavevector, $\mathbf{Q}^\mathrm{eff}_L(T)$, can be detected as a peak in the finite-size spin structure factor $\mathcal{S}_{\mathbf{q}} =  N_\text{spin} ( \langle{\lvert{M_{\mathbf{q}}}\rvert^2}\rangle - \lvert{\langle{M_{\mathbf{q}}}\rangle}\rvert^2)$ and we expect $\lim_{L\to\infty}\mathbf{Q}^\mathrm{eff}_L(T) = \mathbf{Q}(T)$. Indeed, the observed behavior of $\mathbf{Q}^\mathrm{eff}_L(T)$ shows relatively small variance with respect to $L$, supporting this expectation (Fig.~\ref{fig:C_and_Q}). Based on the estimate of $\mathbf{Q}^\mathrm{eff}_L(T)$, we evaluate $\langle{\lvert{M_{\mathbf{Q}^{\mathrm{eff}}_L(T)}}\rvert^2}$ and $\langle{\lvert{M_{\mathbf{Q}^{\mathrm{eff}}_L(T)}}\rvert^4}$, thereby
\begin{align}
    U_{\mathbf{Q}^{\mathrm{eff}}_L}
    = \frac{
        \langle{\lvert{M_{\mathbf{Q}^{\mathrm{eff}}_L(T)}}\rvert^4}\rangle
    }{
        \langle{\lvert{M_{\mathbf{Q}^{\mathrm{eff}}_L(T)}}\rvert^2}\rangle^2
    },
    \label{eq:BR}
\end{align}
which is a Binder parameter defined at $\mathbf{q} = \mathbf{Q}^\mathrm{eff}_L(T)$. As in the usual usage of the Binder parameter,\cite{LandauBinder} we look into the crossing point of $U_{\mathbf{Q}^{\mathrm{eff}}_L(T)}$ for different system sizes to evaluate $\Tsdw$.

Unlike the conventional approach, however, the wavevector $\mathbf{q} = \mathbf{Q}^\mathrm{eff}_L(T)$ associated with the Binder parameter for one system size can be different from one for another size. Furthermore, it can also be $T$-dependent. To analyze the effect of a small deviation of the  wavevector from the true (incommensurate) ordering wavevector $\mathbf{Q}$, we review the standard scaling theory for correlation functions.\cite{Cardy} The two-point SDW correlation function $G_2(\mathbf{r}_1 - \mathbf{r}_2) = \langle{ \SDW(\mathbf{r}_1) \SDW(\mathbf{r}_2) }\rangle$, where $\SDW(\mathbf{r}) \sim \sigma^z(\mathbf{r}) \,e^{-i\mathbf{Q}_c\cdot\mathbf{r}}$ is the coarse-grained SDW order parameter with the momentum $\mathbf{Q}_c \equiv \mathbf{Q}(\Tsdw)$, is expected to have the following transformation under the scaling by a factor $b$,
\begin{align}
    G_2(\mathbf{r},t) 
    \sim b^{-2x^{}_{\mathrm{SDW}}} 
    G_2(
    b^{-1} \mathbf{r}, 
    b^{1/\nu} t),
\end{align}
where $t = (T-\Tsdw)/\Tsdw$, $x_{\mathrm{SDW}}$ is the scaling dimension of the order parameter, and $\nu$ is the critical exponent of correlation length $\xi$. It follows that the correlation function has the universal finite-size scaling form of
\begin{align}
    G_2(\mathbf{r}) 
    \sim \xi^{-2x^{}_{\mathrm{SDW}}} 
    \Psi_2(\xi^{-1}\mathbf{r},\xi^{-1}L),
\end{align}
from which we find
\begin{align}
    \left\langle{\left\lvert{M_\mathbf{q}}\right\rvert^2}\right\rangle
    & \sim \frac{1}{N_\text{spin}^2} \left\langle{
    \left\lvert{
    \int\! \sigma^z(\mathbf{r})\, e^{-i\mathbf{q}\cdot\mathbf{r}}
    d \mathbf{r}\,
    }\right\rvert^2
    }\right\rangle
    \notag\\
    &\sim
    \frac{1}{N_\text{spin}} \int\! G_2(\mathbf{r})\, e^{-i(\mathbf{q}-\mathbf{Q}_c)\cdot\mathbf{r}}
    d \mathbf{r}
    \notag\\
    &\sim
    L^{-2x^{}_{\mathrm{SDW}}}
    \Phi_2(L(\mathbf{q}-\mathbf{Q}_c),\xi^{-1}L).
\end{align}
A similar argument can also be made for the four-point correlation function $G_4(\mathbf{r}_1,\mathbf{r}_2,\mathbf{r}_3,\mathbf{r}_4) = \langle{ \SDW(\mathbf{r}_1) \SDW(\mathbf{r}_2) \SDW(\mathbf{r}_3) \SDW(\mathbf{r}_4) }\rangle$, leading to
\begin{align}
    \left\langle{\left\lvert{M_\mathbf{q}}\right\rvert^4}\right\rangle
    & \sim  
    L^{-4x^{}_{\mathrm{SDW}}}
    \Phi_4(L(\mathbf{q}-\mathbf{Q}_c),\xi^{-1}L),
\end{align}
Thus, near the critical point $T \approx \Tsdw$, we find that the Binder parameter associated with the effective ordering wavevector behaves as
\begin{align}
    U_{\mathbf{Q}_L^{\mathrm{eff}}}
    \sim
    \Phi_U(L(\mathbf{Q}_L^{\mathrm{eff}}-\mathbf{Q}_c),\xi^{-1}L),
    \label{eq:UQ:scaling}
\end{align}
whereas $U_{\mathbf{Q}_L^{\mathrm{eff}}} \to 2$ (corresponding to the Gaussian distribution for a one-component complex order parameter) and $U_{\mathbf{Q}_L^{\mathrm{eff}}} \to 1$ in the high- and low-$T$ limits, respectively. Here, $\Psi_2$, $\Phi_2$, $\Phi_4$, and $\Phi_U$ are  finite-size scaling functions. 

\begin{figure}
\begin{center}
    \includegraphics[
    width=0.9\hsize
    ]{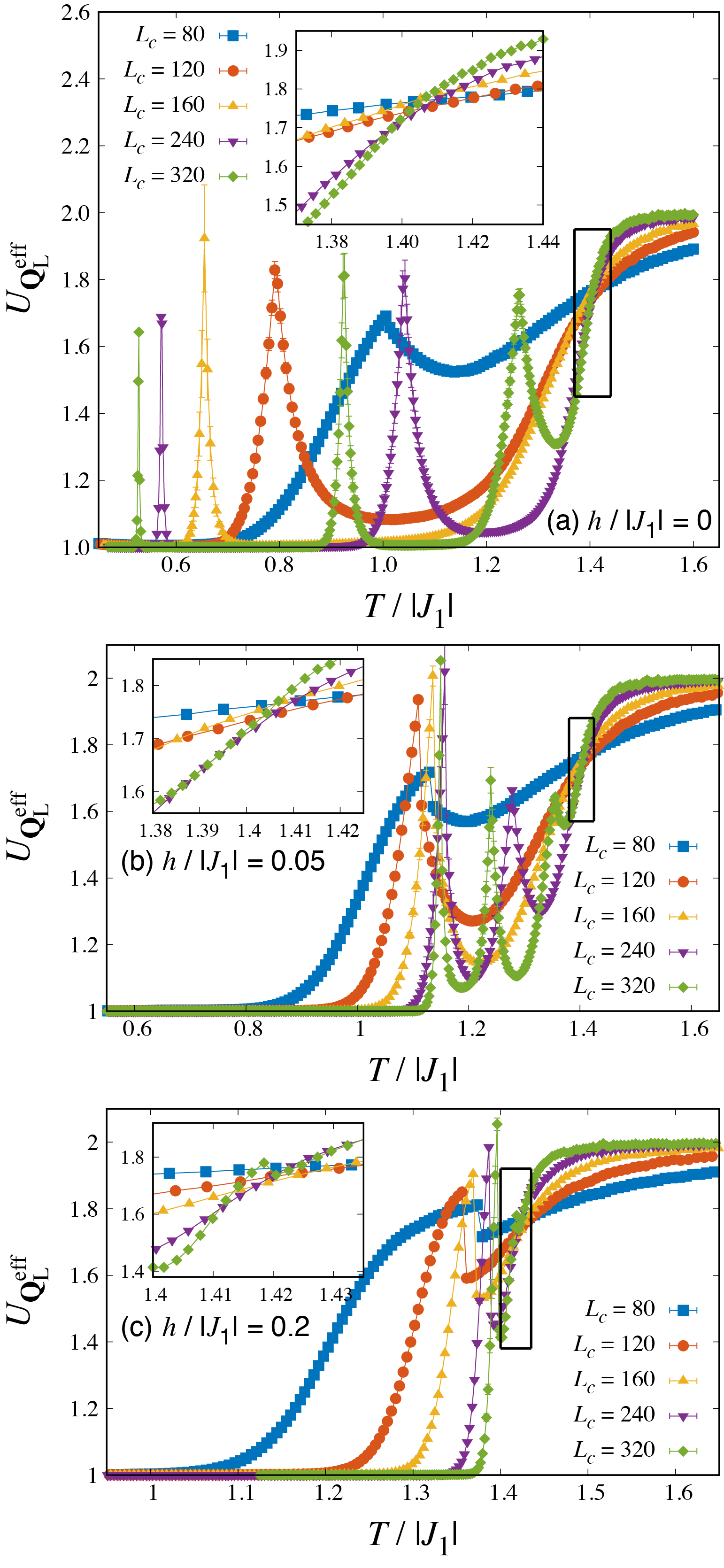}
\end{center}
\caption{
 \label{fig:BinderR}
 Monte Carlo results of the modified Binder parameter $U_{\mathbf{Q}^{\mathrm{eff}}_L}$ [Eq.~\eqref{eq:BR}]
 for (a) $h / \lvert{J_1}\rvert = 0$, (b) $h / \lvert{J_1}\rvert = 0.05$, and (c) $h / \lvert{J_1}\rvert = 0.2$. The insets show enlarged views near $T = \Tsdw$.
}
\end{figure}

\begin{figure*}
\begin{center}
       \includegraphics[
       width=0.95\hsize
       ]{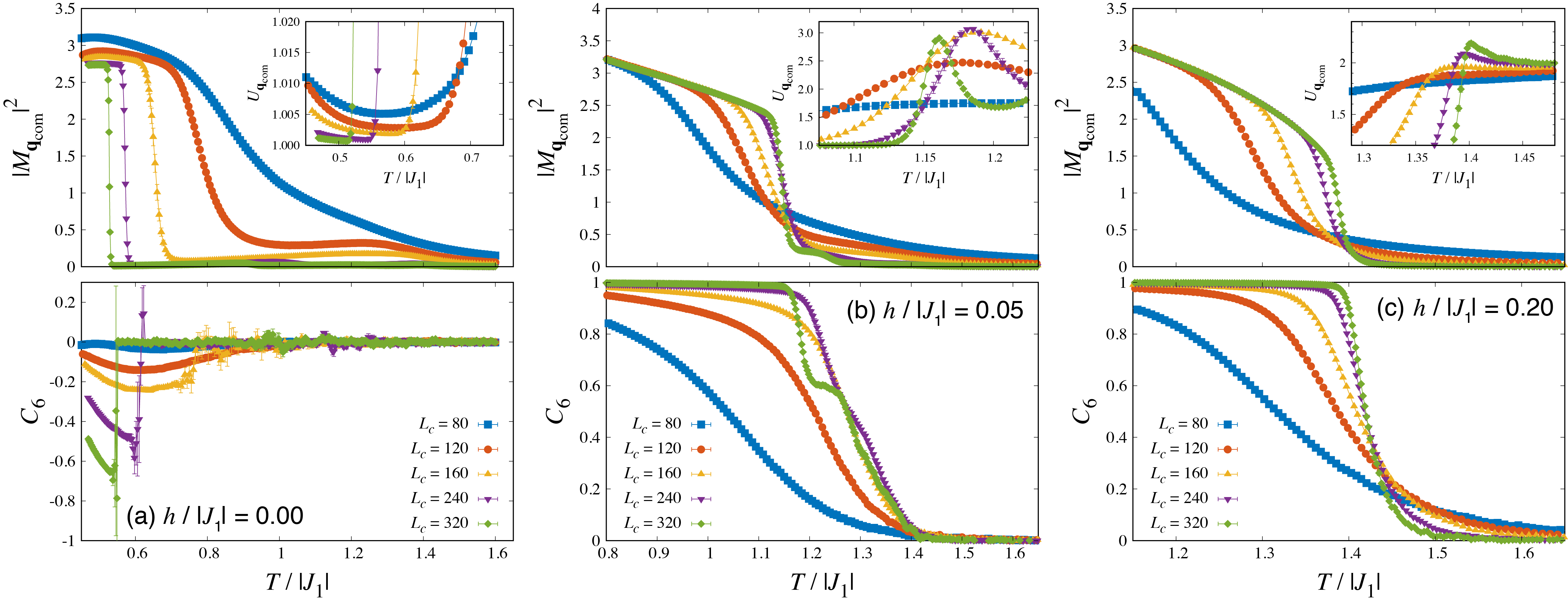}
\end{center}
\caption{
 \label{fig:Mqcom}
 Monte Carlo results of the square of the commensurate order parameter (upper panels) and the $C_6$ indicator to distinguish between the PDA state ($C_6 < 0$) and the FIM state ($C_6 > 0$) (lower panels) for (a) $h / \lvert{J_1}\rvert = 0$, (b) $h / \lvert{J_1}\rvert = 0.05$, and (c) $h / \lvert{J_1}\rvert = 0.2$. The insets show enlarged views of the Binder parameter for the commensurate order parameter near $T = T_{\mathrm{IC-C}}$.
}

\end{figure*}

It is reasonable to assume that these scaling functions are sufficiently isotropic with respect to small $\lvert{\mathbf{Q}_L^{\mathrm{eff}}-\mathbf{Q}_c}\rvert$, with an appropriate rescaling along the principle axes if needed.\cite{Cardy} In fact, $Z_2$ reflection symmetry along the $c$-axis concerning the sign of $Q^\mathrm{eff}_{L,3} - Q_3$ is enough for the following discussion. As a wavevector best approximating the true ordering wavevector for a given system size $L$, we expect $\lvert{\mathbf{Q}_L^{\mathrm{eff}}-\mathbf{Q}_c}\rvert \sim O(L^{-1})$. Since, as usually the case, the universal scaling functions can be expected as a sufficiently slowly-varying function near the critical point $T \approx \Tsdw$, the right hand side of Eq.~\eqref{eq:UQ:scaling} is almost a size-independent constant. Therefore, the Binder parameter at the size-dependent effective ordering wavevectors to exhibit a crossing behavior at around $T = \Tsdw$, as the temperature is lowered from the paramagnetic phase.

In the meantime, to determine $\Tli$ for the lock-in IC-C transition into a commensurate phase in low fields, or $T_c$ for a direct transition into the same phase in high fields, we use the ordinary the Binder parameter $U^{}_{\Qcom} = \langle{\lvert{M_{\Qcom}}\rvert^4}\rangle / \langle{\lvert{M_{\Qcom}}\rvert^2}\rangle^2$ at the corresponding commensurate wavevector $\mathbf{q} = \Qcom = (0,0,2\pi)$.

\subsection{
Ordering wavevector and the phase diagram
}

Below, we discuss the details of the phase diagram for $\kappa = 0.1$ [Fig.~\ref{fig:phase_diagram}(b)], obtained at $T \gtrsim 0.45 \lvert{J_1}\rvert$, focusing on the behavior of the ordering wavevector (Fig.~\ref{fig:C_and_Q}). At low fields, we observe that the ordering wavevector $\mathbf{Q}^\mathrm{eff}_L = (0,0,Q^\mathrm{eff}_{L,3})$ at $T = \Tsdw$ slightly, but clearly, deviates from $\Qcom = (0,0,2\pi)$. Below $\Tsdw$, $\mathbf{Q}^\mathrm{eff}_L$ slowly drifts towards $\Qcom=(0,0,2\pi)$ as further lowering $T$. Roughly speaking, the ordering wavevector changes more rapidly as the magnetic field is increased. The observed step-like behavior of $\mathbf{Q}^\mathrm{eff}_L$ is simply due to the finite-size discretization of the wavevector (e.g., $\Delta Q^{\mathrm{eff}}_{L,3} / (2\pi) = 0.003125$ for $L_c = 320$), which also causes spurious peaks in $C$ at wildly size-dependent temperatures (Fig.~\ref{fig:C_and_Q}). Considering the wide range of the system sizes we investigated, the most natural interpretation is that the change of the wavevector in thermodynamic limit, $\mathbf{Q}(T)=\lim_{L\to\infty}\mathbf{Q}^\mathrm{eff}_L(T)$, is (quasi-)continuous towards $\Qcom$. When $\mathbf{Q}^{\mathrm{eff}}_{L}(T)$ changes from one value to another as a function of $T$ in a finite-size system, a disordering effect appears at large distance due to the mismatch between an ideal wavevector and the system size. The mismatch-induced disordering effect causes a spurious spiky peak in $U_{\mathbf{Q}^{\mathrm{eff}}_L}$ (Fig.~\ref{fig:BinderR}) and interfere with high-precision determination of $\Tsdw$ as the behavior of $U_{\mathbf{Q}^{\mathrm{eff}}_L}$ near the crossing point is affected and becomes less systematic. Therefore, we treat the first crossing points of $U_{\mathbf{Q}^{\mathrm{eff}}_L}$ for $(L_c, 2L_c) = (80,160),\,(120,240),\,\text{and}\,(160,320)$ simply on an equal footing, which nonetheless allows us to determine the transition point of the SDW ordering to the satisfactory precision, e.g., $\Tsdw / \lvert{J_1}\rvert = 1.407(5),\,1.406(1),\,\text{and}\,1.41(2)$ for $h / \lvert{J_1}\rvert = 0, 0.05,\,\text{and}\,0.2$, respectively (Fig.~\ref{fig:BinderR}). The estimated $\Tsdw$ roughly coincides with the highest-$T$ peak in the specific heat (see Fig.~\ref{fig:C_and_Q}). 

At low temperatures, the ordering wavevector is pinned at $\Qcom = (0,0,2\pi)$, corresponding to a three-sublattice ordered phase. At low fields, the low-$T$ phase appears through a lock-in IC-C transition, where the translational symmetry along the $c$ axis, which is broken in the SDW state, is restored. By analyzing the behavior of $U_{\Qcom}$, we find that $\Tli$ increases rapidly with $h$ (see Fig.~\ref{fig:phase_diagram}), e.g., $T_{\mathrm{IC-C}} / \lvert{J_1}\rvert = 0.52(1),\,1.15(3),\,\text{and}\,1.397(3)$ for $h / \lvert{J_1}\rvert = 0, 0.05,\,\text{and}\,0.2$, respectively.  Since $\Tsdw$ is nearly constant in $h$, the SDW phase shrinks rather rapidly with increasing $h$, and above a magnetic field-induced multicritical point $h_\mathrm{LP} \simeq 0.2\lvert{J_1}\rvert \simeq 0.17 h_\mathrm{sat}$, where $h_\mathrm{sat} = 6 (J_2 + J_3)$ is the saturation field, no incommensurate phase exists. The estimated IC-C transition temperatures coincide with the temperatures at which the finite-size ordering wavevector reaches $\mathbf{Q}^{\mathrm{eff}}_L = \Qcom$ at each field, as expected. Also, $\langle{\lvert{M_{\Qcom}}\rvert^2}\rangle$ increases rapidly around the estimated temperatures and becomes almost size-independent at lower temperatures (Fig.~\ref{fig:Mqcom}). While the IC-C transitions studied in the present MC work exhibit typical features of a first-order transition, such as the correction in the finite-size transition temperature varying as $\sim 1/L^3$, they may be a spurious behavior caused by discrete wavevectors in finite size systems, as also found in specific heat as the spurious spiky peaks.

For $T < \Tli$, the two main candidates for the three-sublattice order are the ferrimagnetic (FIM) state and the partially-disordered antiferromagnetic (PDA) state, similar to the case in triangular lattice antiferromagnetic Ising models.\cite{Blankschtein84,Coppersmith85,Isakov03,Heinonen89,Bunker93,Isakov03,Lin14} In the FIM state, each spin chain has a ferromagnetic order and different spin chains takes the three-sublattice $\uparrow\uparrow\downarrow$ structure. In the PDA phase, the spin chains in the first and the second sublattices have a ferromagnetic order with spin-$\uparrow$ and $\downarrow$, respectively, with the spin chains in the third sublattice disordered. A convenient indicator for a finite-size calculation is
\begin{align}
    C_6 = \frac{\langle{M^6_{\Qcom}}\rangle}{ \bigl\langle{\lvert{M_{\Qcom}}\rvert^6}\bigr\rangle},
\end{align}
which takes $C_6 > 0$ ($C_6 < 0$) for the FIM (PDA) state.\cite{Isakov03} For $h = 0$, we confirm the PDA state below $\Tli$ (Fig.~\ref{fig:Mqcom}), though the result should be distinguished from the previous claim of the same state in CCO below $\simeq$\,25\,K.\cite{Kageyama97}
We find no evidence of an additional order-order transition at low $T$ as reported in CCO,\cite{Agrestini11} at least for $T \gtrsim 0.45\lvert{J_1}\rvert$. For $h \ne 0$, we find the FIM state at low $T$ down to $h / \lvert{J_1}\rvert = 0.025$, implying that the observed PDA state is extremely fragile against the magnetic field. The FIM phase yields the 1/3 magnetization plateau.~\cite{Nekrashevic21} At high fields, we find a direct transition into the FIM phase without the intervening SDW phase. We find that the transition is of the first order, which is consistent with the $Z_3$ symmetry breaking in $d=3$, as in the three-state Potts model.\cite{Blote1979,Janke97}

\section{
Ginzburg-Landau theory
\label{Sec:GL}
}

Finally, for a more universal description of the commensurate and incommensurate phases in CCO and similar materials, we consider a GL theory. As will be shown below, the theory is essentially in the same form as that for the ANNNI model.\cite{Aharony1981,Bak80} Interestingly, however, the GL theory for ANNNI-like models in magnetic fields~\cite{Yokoi81} has yet been thoroughly discussed in the literature, despite its experimental relevance.

We use a complex order parameter $\field(\mathbf{r}) \sim e^{-i\Qcom\cdot\mathbf{r}} \sigma^z(\mathbf{r})$, $\Qcom = (0,0,2\pi)$, which can be formally introduced by using the Hubbard-Stratonovich transformation. $\field(\mathbf{r})$ describes the local three-sublattice order in a coarse-grained way, which can be either the FIM order or the PDA order depending on the phase factor. For $h = 0$, we find the following GL Hamiltonian,
\begin{align}
    \GLH_{\,h=0}
    &=
    \int\! \mathrm{d}^3 \mathbf{r}\,
    \Bigl[\,
    \frac{1}{2} \bigl\lvert{
    \left(
    \nabla 
    - i\bm{\epsilon}
    \right)
    \field(\mathbf{r})
    }\bigr\rvert^2
    + t \lvert{ \field(\mathbf{r}) }\rvert^2
    + u^{}_4 \lvert{ \field(\mathbf{r}) }\rvert^4
    \notag\\[-2pt]
    &
    \hspace{60pt}
    + u^{}_6 \lvert{ \field(\mathbf{r}) }\rvert^{6}
    + v^{}_6 \bigl(
    \field(\mathbf{r})^6
    + \field^{\ast}(\mathbf{r})^6
    \bigr)\,
    \Bigr],
    \label{eq:Heff0}
\end{align}    
where $t$, $u^{}_4$, $u^{}_6$, and $v^{}_6$ are GL coefficients and $\bm{\epsilon} = (0,0,\epsilon)$.
A crucial point of the present theory is that the wavevector $\Qcom$ of the local order described by $\field(\mathbf{r})$ is slightly shifted from the minima $\mathbf{q} = \pm\Qmin = \pm(\Qcom + \bm{\epsilon}) $ of the Fourier transform $J(\mathbf{q})$ of the exchange interactions, as briefly mentioned before.
For this reason, $\GLH_{\,h=0}$ includes the vector potential-like (but constant) contribution $- i\bm{\epsilon}$ in the gradient term.
$\GLH_{\,h=0}$ also includes the six-order term ($v_6$) as the leading Umklapp term, the order of which is determined by the size of the magnetic sublattice of the commensurate order and time reversal symmetry; a factor of three comes from $3\Qcom \equiv 0$ and time reversal symmetry requires another factor of two. 

As is clear from the origin, the gradient term acts as adding a momentum $\bm{\epsilon}$ to $\field$, in favor of the three-sublattice long-wavelength SDW state $\langle{ \field(\mathbf{r}) }\rangle \sim \text{(const.)} \times e^{i\bm{\epsilon}\cdot\mathbf{r}}$. Although the SDW state thus appears to benefit from the exchange energy, the sinusoidal modulation made of localized Ising-like moments also requires entropy contribution to the free energy. In contrast, because of the shift of $\Qcom$ from the minima of $J(\mathbf{q})$, the commensurate three-sublattice ordered state, $\langle{ \field(\mathbf{r}) }\rangle \sim \text{const.}$, may appear not to acquire the full energy gain of the exchange interaction. However, the state is quite compatible with the Ising anisotropy. In the GL theory~\eqref{eq:Heff0}, the Umklapp term plays the role of the entropy contribution related with the Ising anisotropy. While the commensurate state may be favored by the Umklapp term by adjusting its constant phase factor, the incommensurate SDW states generally gain no corresponding contribution because of the phase cancellation in the integral over the space. In fact, we could rederive the GL theory in terms of $\SDW(\mathbf{r}) \sim \sigma^z(\mathbf{r}) \,e^{-i\Qmin\cdot\mathbf{r}}$ instead of $\field(\mathbf{r}) \sim \sigma^z(\mathbf{r}) \,e^{-i\Qcom\cdot\mathbf{r}}$, and the result is an ordinary $\SDW^4$-theory for the one-component complex order parameter without Umklapp terms if $\Qmin$ is incommensurate. Hence, the key role in the GL theory~\eqref{eq:Heff0} is played by the competition between the gradient and the Umklapp terms favoring incommensuration and commensuration, respectively. Thus, although somewhat different in appearance, the Hamiltonian of this system~\eqref{eq:H} realizes essentially the same kind of situation as the classic ANNNI model.\cite{Bak80}

At $T = \Tsdw$, critical fluctuations renders $\field(\mathbf{r})$ nonzero with the additional momentum $\bm{\epsilon}$, resulting in the SDW state with the ordering wavevector $\mathbf{Q} = \Qcom + \bm{\epsilon} = \Qmin$. In other words, we expect a condensation of the softest mode $\SDW(\mathbf{r}) \sim \sigma^z(\mathbf{r}) \,e^{-i\Qmin\cdot\mathbf{r}}$ rather than $\field(\mathbf{r}) \sim \sigma^z(\mathbf{r}) \,e^{-i\Qcom\cdot\mathbf{r}}$. Since the Umklapp $v^{}_6$ term has no effect for the incommensurate SDW states, the transition, which breaks translation symmetry along the $c$ axis, will be in the 3D \textit{XY} universality class (emergent U(1) symmetry). The observed main peak of the specific heat, which exhibits a sign of smearing [Fig.~\ref{fig:C_and_Q}(a)], is consistent with the negative exponent $\alpha = -0.0146(8) < 0$ for the \textit{XY} universality class.\cite{Campostrini2001}

For $T < \Tsdw$, the competition between the gradient term and the Umklapp term sets in, which affects the phase factor of $\langle{\field(\mathbf{r})}\rangle$, thereby $\mathbf{Q}(T< \Tsdw)$. To see this, we may write $\field(\mathbf{r}) = A(\mathbf{r}) e^{i\theta(\mathbf{r})}$ and apply a mean-field decoupling for the massive amplitude fluctuation (``Higgs'')  mode $\delta A(\mathbf{r}) = A(\mathbf{r}) - \langle{A}\rangle$ and the phase mode $\theta(\mathbf{r})$.\cite{Bak80} The result is the following sine-Gordon model for $\theta(\mathbf{r})$,
\begin{align}
   \GLH'_{\,h=0,\,\theta} = 
   \langle{A}\rangle^2\!
   \int\! \mathrm{d}^3 \mathbf{r}\,
   \Bigl[
   \frac{1}{2}   
   \left(\nabla\theta(\mathbf{r}) - 
   \bm{\epsilon}
   \right)^2
   + 2 \langle{A}\rangle^4 v^{}_6 \cos6\theta(\mathbf{r})
   \Bigr].
   \label{eq:Heff:phi:1}
\end{align}
The gradient term tends to drift the phase, which can lead to a plethora of soliton lattices through the competition against the cosine term,\cite{Bak80,Chaikin-Lubensky1995} in good agreement with our mean field and MC studies. In the meantime, because the order parameter amplitude $\langle{A}\rangle$ increases as $T$ is lowered below $\Tsdw$, the strength of the cosine term is enhanced. Consequently, the model at low $T$ is expected to undergo a lock-in transition eventually. For $v^{}_6 > 0$ ($v^{}_6 < 0$), the phase is locked-in at $\theta = 2n\pi/6$ [$\theta = (2n+1)\pi/6$] with an integer $0 \le n < 6$, corresponding to the FIM ($\uparrow\uparrow\downarrow$ or $\downarrow\downarrow\uparrow$) state and the PDA state,\cite{Blankschtein84,Coppersmith85,Isakov03,Heinonen89,Bunker93,Lin14} respectively. Although our mean field calculation implies $v_6 > 0$ while our unbiased MC simulation suggests $v_6 < 0$ in zero field, the subtle discrepancy does not require a serious attention as $v_6$ is generated through fluctuations.

For $h \ne 0$, the uniform $\mathbf{q} = 0$ component $\mzero(\mathbf{r})$ is allowed by symmetry and may be induced by the magnetic field. Consequently, in addition to $v_6$, a \emph{lower-order} Umklapp term appears in the GL Hamiltonian. We find
\begin{align}
    \Delta\GLH_{\,0\mathrm{-}\Qcom}
    &\simeq
    \int \mathrm{d}^3 \mathbf{r}\,
    w^{}_4 \mzero(\mathbf{r}) 
    \left(
    \field(\mathbf{r})^3 + \field^\ast(\mathbf{r})^3
    \right)
    \label{eq:H:h:0-Q}
\end{align}
as the leading-order contribution with the new coupling constant $w^{}_4$. The total effective GL Hamiltonian for $h \ne 0$ is
\begin{align}
\GLH_{\,h \ne 0,\,\field,\,\mzero} = \GLH_{\,h=0,\,\field} + \GLH_{\,\mathbf{q}=0,\,m} +
\Delta\GLH_{\,0\mathrm{-}\Qcom},
\end{align}
where 
$
\GLH_{\,\mathbf{q}=0,\,m} = \int\! \mathrm{d}^3 \mathbf{r}\,
\bigl[
    \frac{1}{2} \bigl(
    c\nabla \mzero(\mathbf{r})
    \bigr)^2    
    + \mu^2 \mzero(\mathbf{r})^2
    - h \mzero(\mathbf{r})
\bigr]
$
is the noninteracting part for $\mzero(\mathbf{r})$ with $c>0$ and $\mu^2 > 0$ being the gapped spin wave parameters near $\mathbf{q} = 0$. By a similar mean field decoupling as in the $h = 0$ case, we find a new term in the sine-Gordon model,
\begin{align}
    \Delta\GLH'_{\,0\mathrm{-}\Qcom,\,\theta}
    =
    2 w^{}_4 \langle{\mzero}\rangle
    \langle{A}\rangle^3
    \!
    \int \mathrm{d}^3 \mathbf{r}\,
    \cos3\theta(\mathbf{r}).
   \label{eq:Heff:phi:2}    
\end{align}

The field-induced Umklapp term has the following consequences. Firstly, because of the reduced symmetry in the $\theta$ space, only a subset of the FIM states is favored by $\Delta\GLH'_{\,0\mathrm{-}\Qcom,\,\theta}$ among the three-sublattice ordered states. Depending on the sign of $w_4$, the favored states are  $\uparrow\uparrow\downarrow$ or $\downarrow\downarrow\uparrow$, each of which is three-fold degenerate, though $\uparrow\uparrow\downarrow$ is naturally anticipated for $h > 0$. Secondly, as the prefactor is $\propto \langle{A}\rangle^3$, as opposed to $\propto \langle{A}\rangle^6$ in zero field, the strength of the field-induced Umklapp term is expected to grow faster for $T < \Tsdw$. Moreover, since the prefactor is $\propto \langle{\mzero}\rangle$, we expect that this trend is further enhanced for larger $h$. Therefore, the region of the incommensurate SDW microphases is expected to become narrower for larger $h$, in excellent agreement with our MC results (Fig.~\ref{fig:phase_diagram}). Thirdly, considering that our MC simulation shows the PDA state at $h = 0$ below $\Tli$, the zero- and the field-induced Umklapp terms ($\sim \cos6\theta,\, \cos3\theta$, respectively) must compete against each other in the present system. The competition opens a possibility of a kind of mixed phase below $\Tli$, though our MC results shows no evidence down to an extremely low field, $h / \lvert{J_1}\rvert = 0.025$. Finally, near $T = \Tsdw$, the present GL theory suggests the condensation of the softest mode $\SDW(\mathbf{r}) \sim \sigma^z(\mathbf{r}) \,e^{-i\Qmin\cdot\mathbf{r}}$ rather than $\field(\mathbf{r}) \sim \sigma^z(\mathbf{r}) \,e^{-i\Qcom\cdot\mathbf{r}}$, as in the case of zero magnetic field. 
Hence, we expect the emergent U(1) symmetry at $T = \Tsdw$ also for $h \ne 0$, because Umklapp terms disappear from the GL theory in terms of the incommensurate critical mode $\SDW$. 
The theory thus predicts $\mathbf{Q}(T = \Tsdw) = \Qmin$ for any magnetic field, which is indeed consistent with our MC results at low magnetic fields. However, the simulation closer to the magnetic field-induced multicritical point $h_\mathrm{LP} \simeq 0.2\lvert{J_1}\rvert$ might point to a deviation from this behavior, suggesting that $\mathbf{Q}(T = \Tsdw)$ approaches towards $\Qcom$ for the larger systems we investigated  [Fig.~\ref{fig:C_and_Q}(c)]. This observation may be an indication that the multicritical point is a Lifshitz point induced by a magnetic field.

\section{
Summary and outlook
\label{sec:summary}}

To summarize, we presented the magnetic phase diagram in equilibrium of the 3D spin model for the frustrated quasi-one-dimensional triangular Ising antiferromagnet Ca$_3$Co$_2$O$_6$. We identified the region of incommensurate SDW microphases in a magnetic field (Fig.~\ref{fig:phase_diagram}). We found the deformation of SDW microphases as a function of $T$, characterized by the temperature dependence of the ordering wavevector $\mathbf{Q}(T)$, occurring much more rapidly in a magnetic field than in zero field (Fig.~\ref{fig:C_and_Q}). The deformation eventually leads to the IC-C transition into the PDA (FIM) state for $h = 0$ ($h \ne 0$). Between the PDA and the FIM phases, there may be a mixed phase in an extremely low-field regime, though not confirmed in this work. 
The GL theory we derived includes different symmetry-allowed Umklapp terms for $h = 0$ and $h \ne 0$. The GL theory allowed for further deriving an effective sine-Gordon model that provides a qualitative explanation of the observed magnetic field-induced deformation of the SDW microphases. Moreover, these effective theories demonstrate that the present system can be seen as an incarnation of the classic ANNNI model,\cite{Bak80} despite different appearance of the lattice structure and the complicated network of the exchange interactions. 

Finally, we discuss the relation between the theoretical phase diagram in this work and the previous experiments. As mentioned in Introduction, the material is known for the intriguing combination of the slow relaxation phenomena and the long-wavelength SDW order.
As the recent field-cooling study suggested,\cite{Nekrashevic21} the slow relaxation at low temperature may be greatly influenced by the cooling process passing through the low-field SDW phase at intermediate temperature. 
To verify the conjectured relation experimentally, the challenge is that experiments under equilibrium conditions are known to be notoriously difficult for Ca$_3$Co$_2$O$_6$. 
For example, a resonant X-ray experiments reported a field-induced IC-C transition at 5~K,\cite{Mazzoli09} which is unfortunately most likely non-equilibrium because the temperature is too low. However, at intermediate temperatures above $\Tsf$, there are some experiments that seem to capture the desired physics of the field-induced deformation of the SDW phase and the IC-C transition. For example, a $\mu$SR measurements at 20~K reported a magnetic field-induced phase transition at around 0.4~T.\cite{Takeshita2007} Although the original interpretation of the result suggested a PDA-FIM transition, the obtained phase boundary is very similar to the SDW-FIM transition line shown in the present work. Since the SDW phase was not confirmed back then, it is quite possible that the anomaly in the $\mu$SR experiment is the sign of the IC-C transition induced by the magnetic field. We also note that a similar phase diagram was obtained also by weak anomaly in the magnetic entropy change.\cite{Lampen14} We thus believe that further experiments studying the field-induced IC-C transition in Ca$_3$Co$_2$O$_6$, such as neutron scattering and other spectroscopies focusing on the low-field regime, will be very promising, especially when combined with the recently proposed field-cooling protocol.\cite{Nekrashevic21} Such experiments may provide further insights in the peculiar slow dynamics and out-of-equilibrium behaviors in Ca$_3$Co$_2$O$_6$ and related materials.

\begin{acknowledgments}
The author is grateful to valuable discussions with Vivien Zapf, Xiaxin Ding, Ivan Nekrashevich, and Cristian Batista. The author acknowledges the support by the NSFC (Nos.~11950410507, 12074246, and U2032213) and MOST (No.~2016YFA0300501) research programs.
\end{acknowledgments}

\bibliographystyle{apsrev4-1} 
\bibliography{refs}

\end{document}